\documentclass{PoS}

\title{Higgs Physics at future Linear Colliders -- A Case for precise Vertexing}

\ShortTitle{Higgs Physics at future Linear Colliders -- A Case for precise Vertexing}

\author{\speaker{Frank Simon}%
      \\
       Max-Planck-Institut f\"ur Physik\\
       E-mail: \email{fsimon@mpp.mpg.de}}

\abstract{The discovery of a Higgs boson by the experiments at the LHC marks a major breakthrough in particle physics, with far-reaching consequences for our understanding of the fundamental principles of our Universe. To fully explore this unique particle, experiments at high-energy electron-positron colliders are being planned, providing substantial added benefit over the capabilities of the LHC alone, such as model-independent measurements of couplings, constraints on invisible decays and precise measurements of the self-coupling. This contribution summarizes the Higgs physics program at such future facilities, highlighting in particular also the role of precise vertexing in achieving the ambitious goals of these experiments.}

\FullConference{22nd International Workshop on Vertex Detectors\\
		 September 15th-20th, 2013\\
		 Lake Starnberg, Germany}

\begin{document}

\section{Introduction}

The  discovery of a Higgs boson at the Large Hadron Collider \cite{:2012gk, :2012gu} has substantial consequences for our understanding of the structure of matter, calling for a detailed investigation of the properties and interactions of this particle. It is crucial to precisely establish its mass and its quantum numbers, its coupling to fermions, bosons and its self-coupling, and to determine if it is the single, fundamental scalar predicted by the Standard Model or if it is a part of a more extended Higgs sector, or a composite state bound by so-far unknown interactions. The mass of around 125 GeV provides for a large variety of final states with sufficiently large branching fractions, allowing detailed investigations of the mass dependence of the Higgs couplings at colliders. 

Over the coming years, the LHC is expected to provide decisive answers on some of the questions outlined above. However, a full exploration of this new sector of particle physics will not be possible with the LHC alone. An energy-frontier $e^+e^-$ collider operated at several different energies from 250 GeV up to the TeV region provides substantial additional precision in most areas of Higgs physics and enables measurements not possible at hadron colliders. Most notably, such a facility is capable of fully model-independent measurements of the couplings to bosons and fermions and a measurement of the total width. This increased precision will allow accurately identifying possible non-Standard Model Higgs sectors, which may manifest themselves in deviations of the couplings, which are expected at the percent level for gauge bosons and at the few 10\% level for fermions in typical two-Higgs-doublet models \cite{Gupta:2012mi}.

The full energy reach of such a Higgs program in $e^+e^-$ collisions can only be covered by linear colliders, extending from the threshold of HZ production to the TeV and possibly the multi-TeV region. Two concepts for such a collider are currently being developed within the Linear Collider Collaboration, based on different acceleration schemes which result in different energy reaches. The International Linear Collider (ILC) \cite{Behnke:2013xla} is based on superconducting RF structures while the Compact Linear Collider (CLIC) \cite{Lebrun:2012hj} uses normal-conducting two-beam acceleration technology. For the ILC, the technical design report has recently been completed, while for CLIC the conceptual design report was delivered in 2012, with the technical design phase still ongoing until 2018. The ILC is planned as a 500 GeV machine, starting at 250 GeV with several steps up to the nominal design energy, and an upgrade to 1 TeV. The higher acceleration gradient of the CLIC technology provides a reach up to 3 TeV, with the machine foreseen to be implemented in stages to maximize the physics potential. 

These proceedings present an updated and extended version of \cite{Simon:2012ik}. In the following, the Higgs physics program at a Linear Collider at various energies will be outlined, naturally divided by the physics accessible in the region below a center-of-mass energy of 500 GeV and the region of 500 GeV and above. The discussion is independent of the detailed choice of the machine technology or the detector concept. The results presented here are based on detailed detector simulations performed in the context of the ILC and CLIC physics and detector studies, and are taken from recent reports \cite{Behnke:2013lya, Asner:2013psa, Abramowicz:2013tzc}, partially with recent updates for the currently ongoing P5 process in the United States. In addition, the detector requirements, in particular concerning the vertex detector, originating from this physics program are summarized.

\section{Higgs Production at Linear Colliders}

\begin{figure}
\centering
\includegraphics[height=0.2\textwidth]{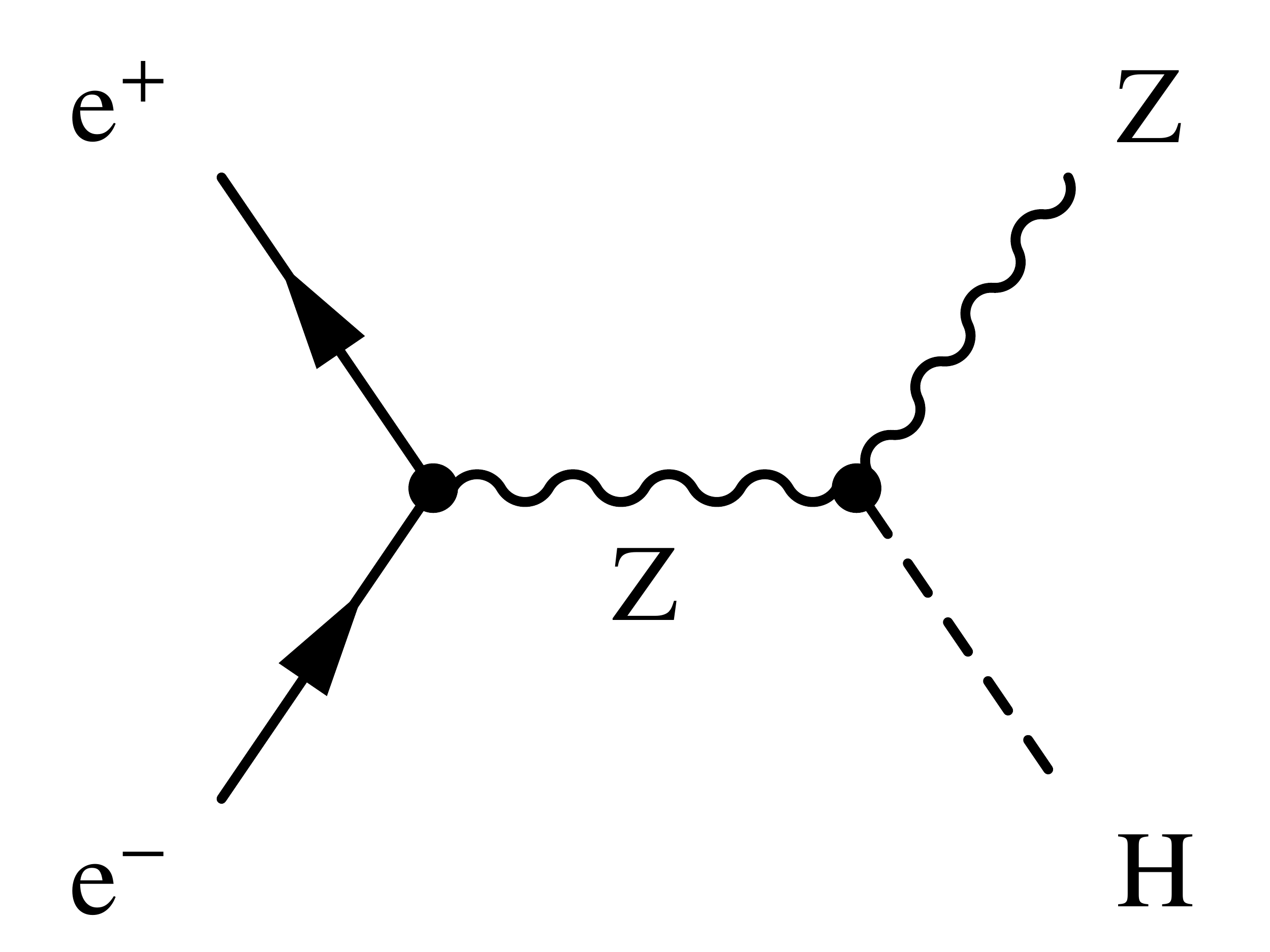}
\hspace{0.1\textwidth}
\includegraphics[height=0.22\textwidth]{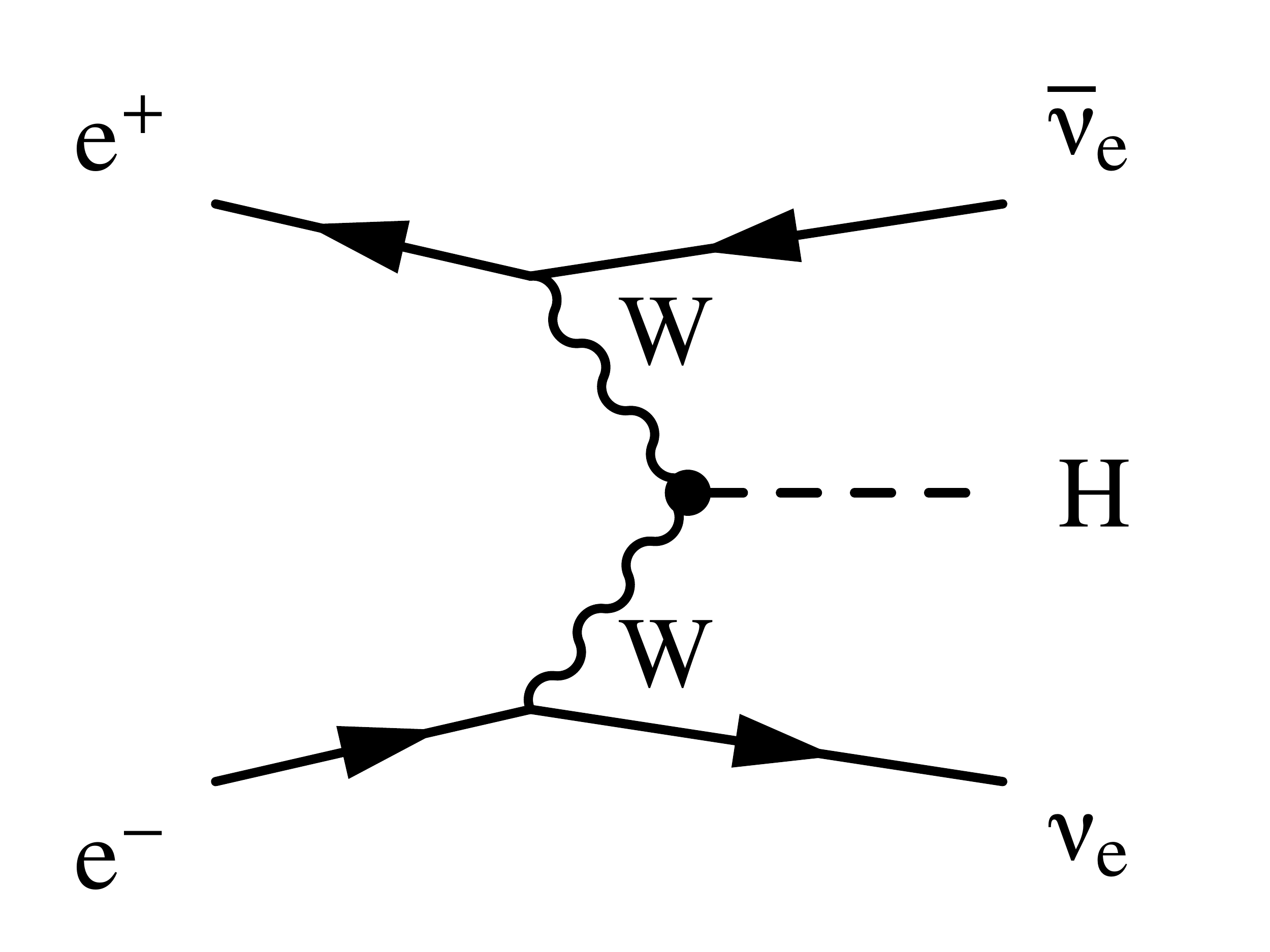}\\
\vspace{0.02\textwidth}
\includegraphics[height=0.2\textwidth]{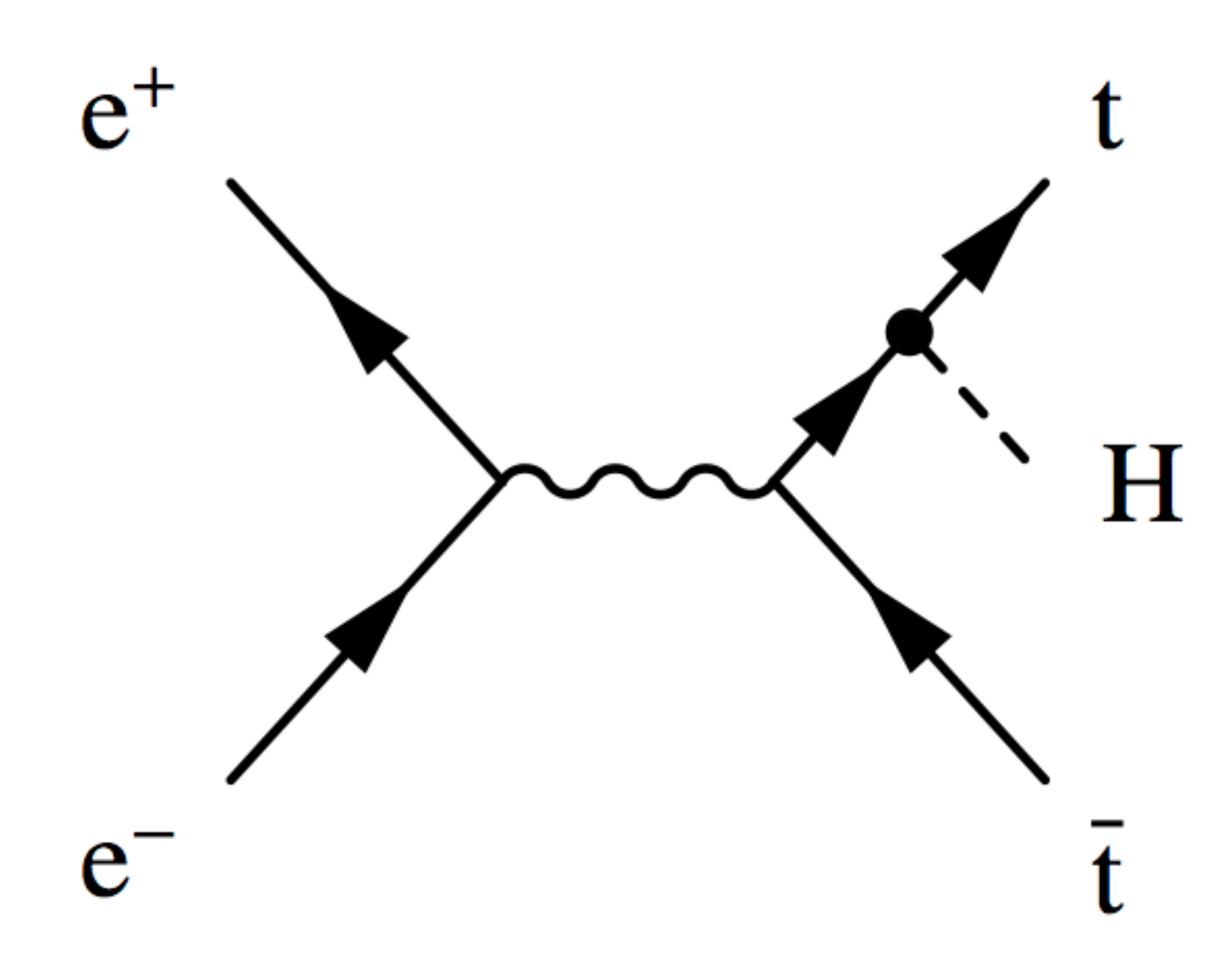}
\hspace{0.05\textwidth}
\includegraphics[height=0.2\textwidth]{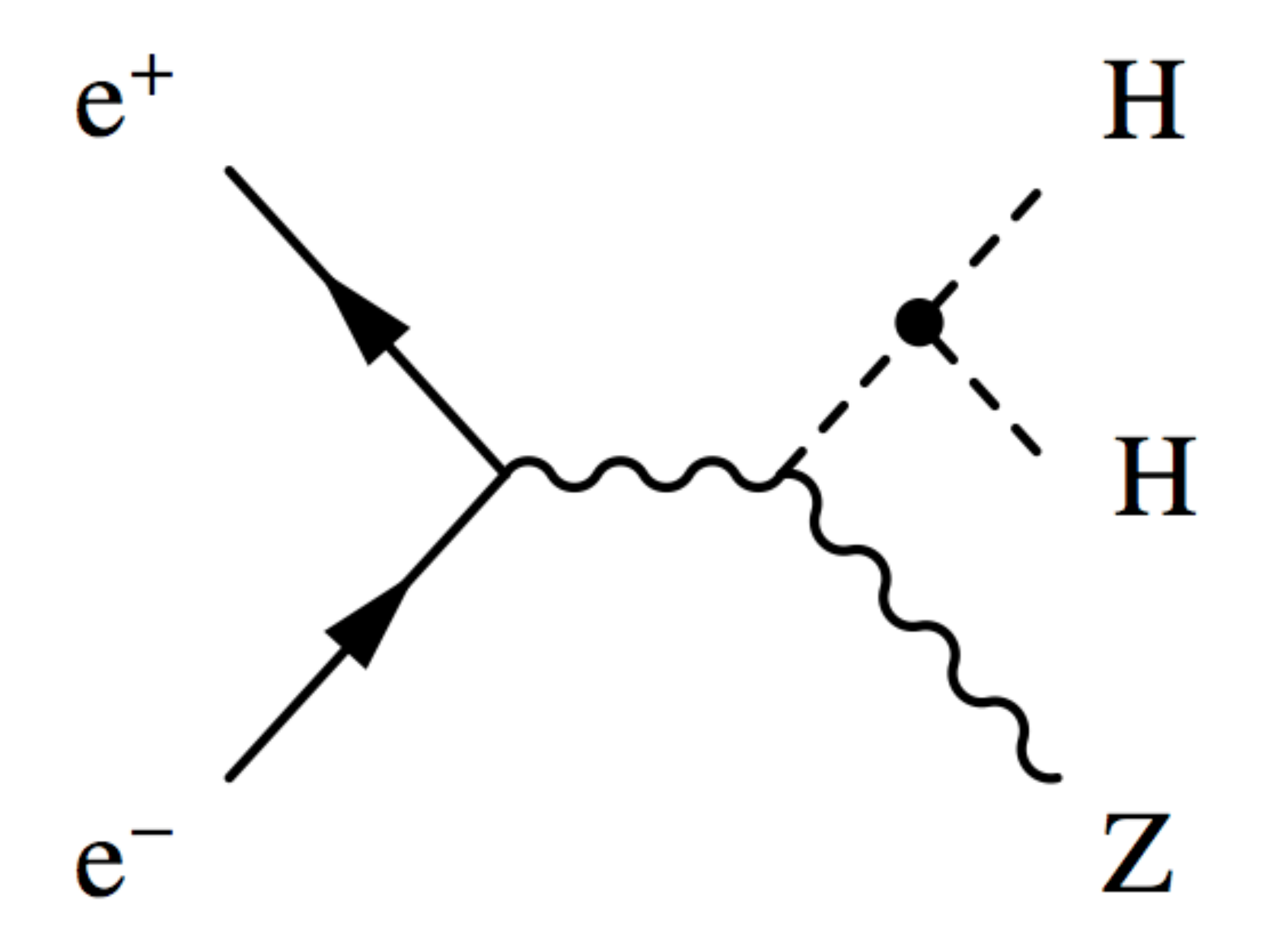}
\hspace{0.05\textwidth}
\includegraphics[height=0.2\textwidth]{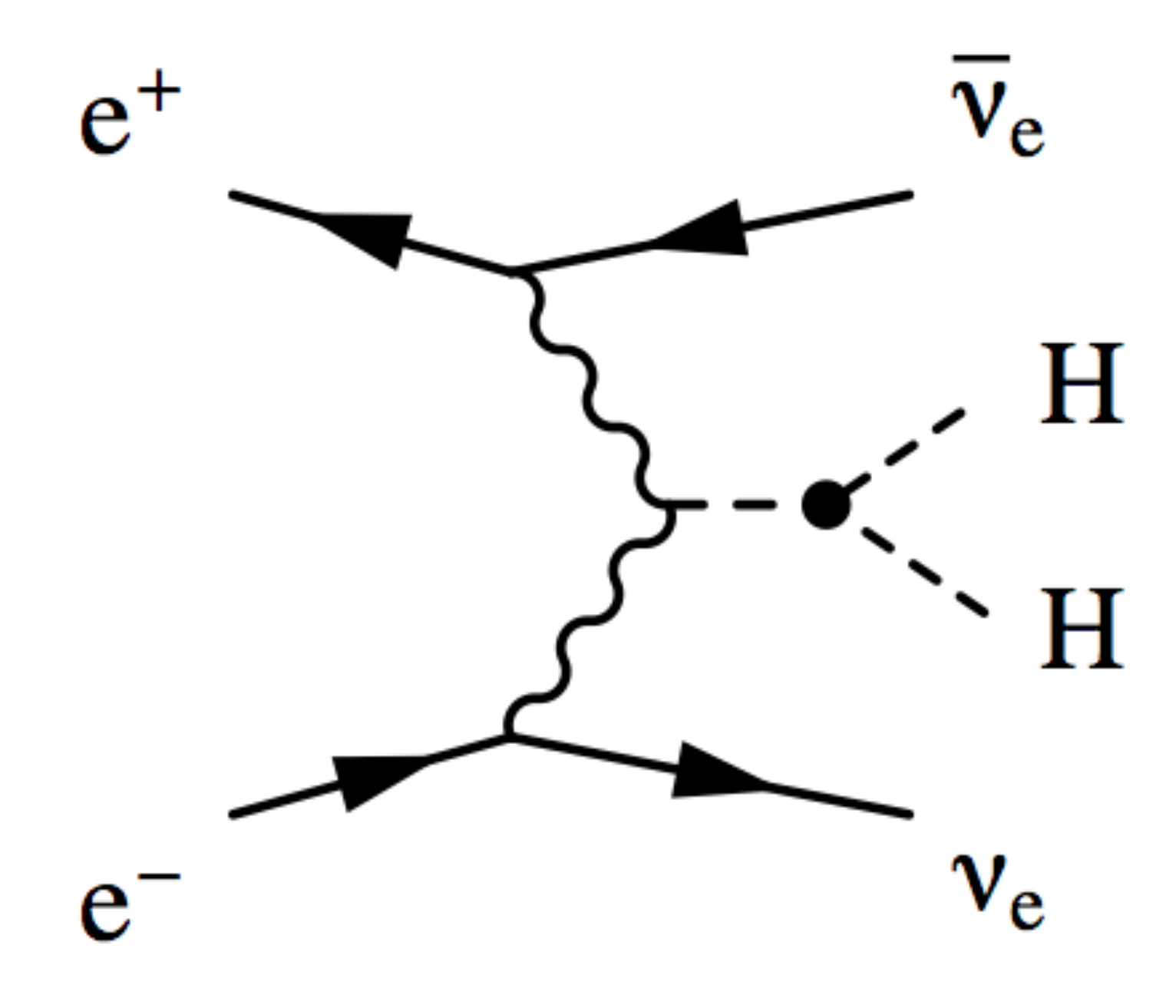}

\caption{Higgs production in $e^+e^-$ annihilation. Top Row: The two leading processes, Higgs-strahlung and vector boson fusion. Bottom row: Top-Higgs production and double Higgs production in the strahlungs- and the fusion process.}
\label{fig:HZ}
\end{figure}

\begin{figure}
\centering
\includegraphics[width=0.498\textwidth]{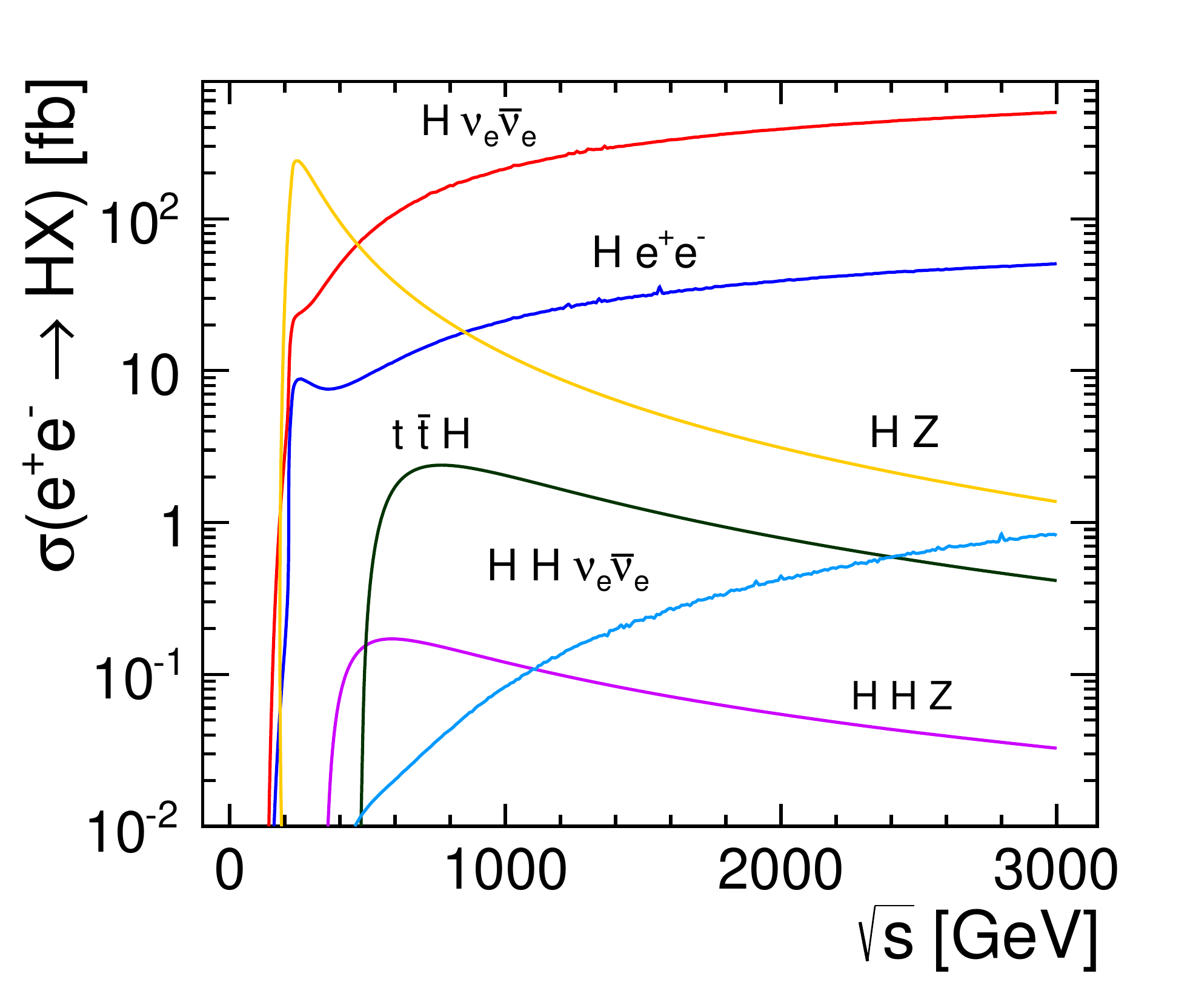}
\hfill
\includegraphics[width=0.441\textwidth]{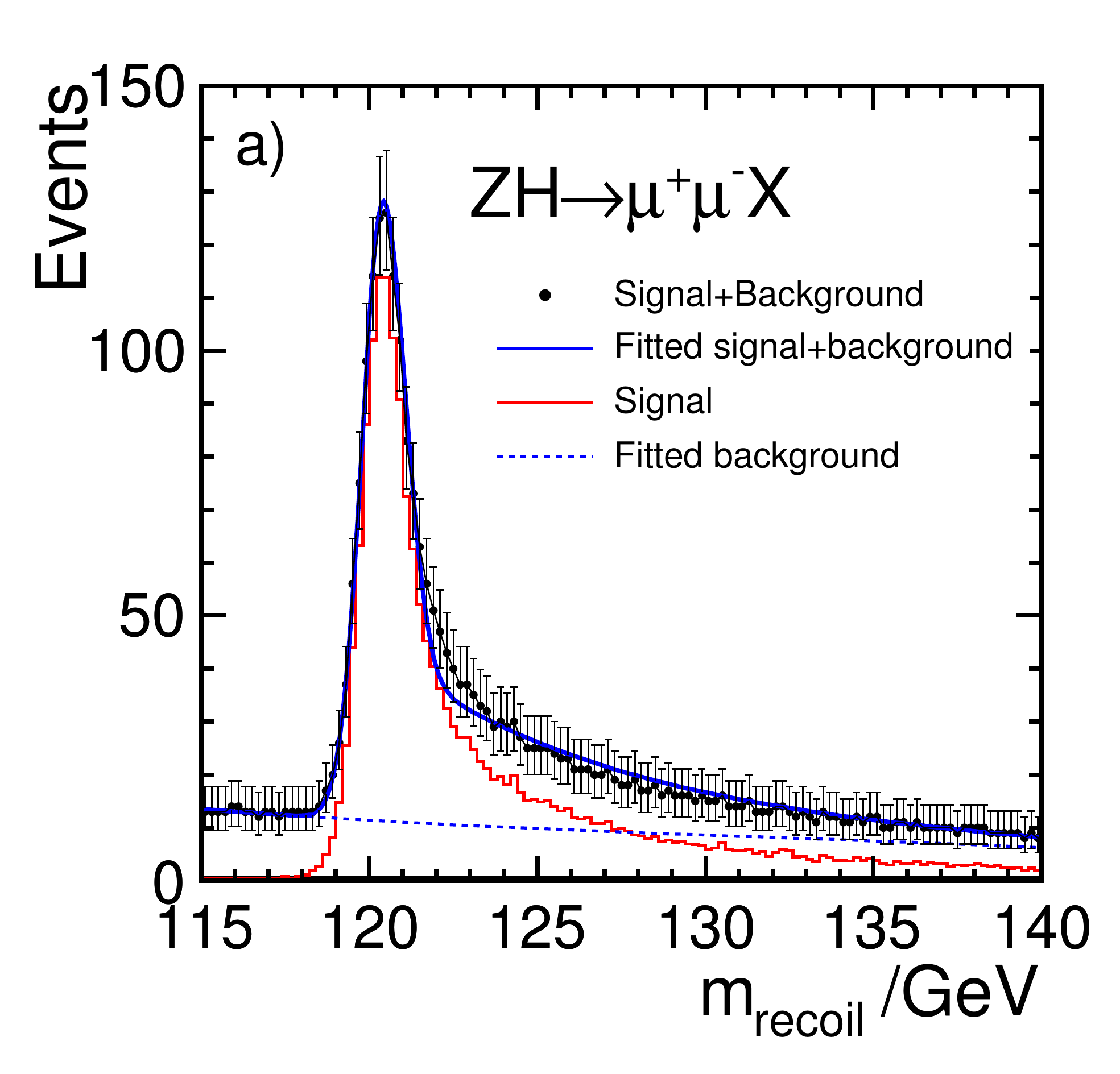}
\caption{Left: Tree-level cross sections for various Higgs processes as a function of $e^+e^-$ collision energy for unpolarized beams. Right: Recoil mass distribution for the process $e^+e^-\rightarrow \mathrm{ZH} \rightarrow \mu^+\mu^-X$ together with non-Higgs background at a center-of-mass energy of 250 GeV with an integrated luminosity of 250 fb$^{-1}$ (simulated with $m_\mathrm{H}$ = 120 GeV). The error bars show the expected statistical uncertainty at each point. Figure taken from \cite{Baer:2013cma}.}
\label{fig:CrossSections}
\end{figure}

Higgs bosons are produced in $e^+e^-$ collisions by two main processes, the s-channel Higgs-strahlung process where the Higgs boson is radiated off a Z boson, and the t-channel W boson fusion process, as shown in the top row of Figure \ref{fig:HZ}. LHC has provided evidence for the coupling of the new boson to both W and Z bosons \cite{:2012gk, :2012gu}, confirming that these production mechanisms are accessible. At low center-of-mass energies, the process $e^+e^-\rightarrow \mathrm{ZH}$ dominates, with a cross-section maximum at approximately 250 GeV. Since this cross-section falls rapidly with increasing energy, while the vector boson fusion cross-section increases logarithmically with energy, the \mbox{$e^+e^-\rightarrow \mathrm{H}\nu\nu$} process dominates at energies above approximately 450 GeV for unpolarized beams, as illustrated in \mbox{Figure \ref{fig:CrossSections} {\it left}}. Depending on energy and integrated luminosity, 10$^5$ to 10$^6$ Higgs bosons are expected to be produced at each energy stage, with the highest numbers reached at CLIC at 3 TeV due to the high production cross-section combined with the high instantaneous luminosity of a multi-TeV linear collider.

At energies of 500 GeV and above, additional processes, such as $e^+e^-\rightarrow t\bar{t}\mathrm{H}$,  $e^+e^-\rightarrow\mathrm{ZHH}$ and $e^+e^-\rightarrow \mathrm{HH}\nu\nu$ become accessible, illustrated in the bottom row of  Figure \ref{fig:HZ}. As shown in Figure \ref{fig:CrossSections} {\it left}, these processes have substantially lower cross-sections than the leading production modes, and require high integrated luminosities for precise investigations.

In general, the use of polarized beams, which is possible and foreseen at linear colliders, provides a substantial increase of the signal cross sections in particular in WW fusion processes,  further enhancing the potential for Higgs physics. For instance, the signal and background cross section of double Higgs production are increased by a factor of two with electron (80\%) and positron (30\%)  polarization at 3 TeV.

\section{Measurements below 500 GeV}

The relatively clean environment and the well-defined collision energy at a linear collider together with the precise momentum resolution provided by the detectors allow to make a model-independent measurement of the HZZ coupling in the Higgs-strahlung process. By reconstructing only the decay products of the Z boson, the cross-section of the ZH process can be determined through the recoil mass spectrum. The highest precision is achieved for Z$\rightarrow\mu^+\mu^-$, as shown for the ILD detector at 250 GeV with an integrated luminosity of 250 fb$^{-1}$ in Figure \ref{fig:CrossSections} {\it right}. The distribution peaks at the Higgs mass, with the tail to higher masses due to the combined effects of initial state radiation and beamstrahlung on the center-of-mass energy of the $e^+e^-$ annihilation.  

At energies of 250 GeV and 350 GeV, a precision on the level of 2\% on the cross-section of the Higgs-strahlung process, translating to a precision of 1\% on the coupling $g_\mathrm{HZZ}$, can be achieved when combining several decay modes of the Z, possibly including hadronic decays. Typically, the highest precision is achieved at 250 GeV, due to the higher resolution for lower-momentum muons and due to the higher cross-section, which is somewhat offset by the increased luminosity available at 350 GeV.  With the same process, possible invisible decays of the Higgs boson can be constrained to below the 1\% level.

The explicit reconstruction of the Higgs boson in addition to the recoiling Z boson permits the measurement of the branching fractions of the decays into $b$ and $c$ quarks, $\tau$, WW$^*$, ZZ$^*$ and gluons. 
For these measurements precise flavor tagging to separate $b$ and $c$ jets as well as light jets is crucial, in addition to excellent particle flow performance for the efficient identification of $\tau$ leptons and for the reconstruction of hadronic W and Z decays. Due to more favorable background conditions and improved flavor tagging performance arising from higher boosts, a slightly higher precision is typically achieved at 350 GeV compared to 250 GeV for these measurements. The measurements of $\sigma \times \mathrm{BR}$ for the individual final states, together with the model-independent measurement of $g_\mathrm{HZZ}$  and the total width gives access to the couplings in a model-independent manner. The measured $\sigma \times \mathrm{BR}$ for a given process depends on
\begin{equation}
\sigma(\mathrm{HZ}, \mathrm{H}\nu\nu)\, \times \mathrm{BR}(\mathrm{XX}) \propto g_\mathrm{HVV}\, g_\mathrm{HXX} / \Gamma_\mathrm{H}, \nonumber
\end{equation}
 where $g_\mathrm{HVV}$ is the relevant coupling in the production process, with V = Z for the Higgs-strahlung process and V = W for WW fusion, $g_\mathrm{HXX}$  is the coupling to the final-state particles and $\Gamma_\mathrm{H}$ is the total width of the Higgs. 
 
 The width of the Higgs is too small to be measured directly from the line shape, but it can be extracted by measuring both production and decay via the same particles, using $\Gamma_\mathrm{H} \propto g^2_\mathrm{HXX} / \mathrm{BR}(\mathrm{H} \rightarrow \mathrm{XX})$. In ZH production alone, this requires the measurement of the branching fraction of $\mathrm{H}\rightarrow\mathrm{ZZ^{*}}$. Due to the low branching fraction of this decay, a precise measurement of the total width is only possible when combining Higgs-strahlung and WW fusion measurements, which are accessible at energies around 350 GeV, exploiting $\Gamma_\mathrm{H} \propto \sigma(\mathrm{WW\ fusion}) / \mathrm{BR}(\mathrm{H}\rightarrow\mathrm{WW}^*)$. Here, the total WW fusion cross section is determined from the model-independent measurement of the ZH cross-section and the two measurements of $\sigma \times \mathrm{BR}$ of $\mathrm{H}\rightarrow b\bar{b}$ in ZH and WW fusion, respectively.

 From global fits to all measurements of $\sigma$ and $\sigma \times \mathrm{BR}$, the different couplings and the total width can be determined. For a model-independent extraction, the expected precision ranges from the 1\% -- 2\% level  for $g_\mathrm{HZZ}$, $g_\mathrm{HWW}$  and $g_{\mathrm{H}b\bar{b}}$ to 2\% to 3\% for the other couplings with the exception of the rare decay $\mathrm{H} \rightarrow \mu^+\mu^-$. Also the decay  $\mathrm{H} \rightarrow gg$ will be measured with a precision of $\sigma \times \mathrm{BR}$ of slightly above 2\%. While this can not be directly transformed to a coupling, it provides model-dependent sensitivity to the coupling to the top quark through loop contributions. The total width will be determined with a precision of approximately 5\% with the same fits.  If the model-dependence is given up and a fit strategy which is also applicable to hadron collider data is used \cite{LHCHiggsCrossSectionWorkingGroup:2012nn}, the uncertainties on the coupling deviations from their Standard Model expectation reduces substantially below 1\% for most couplings, in particular when the high statistics available at high collision energies in the TeV and multi-TeV range are considered. 

In addition to the measurement of the couplings, the measurement of the Higgs mass is possible with a precision around 30 MeV or better, both by direct reconstruction of the full final state, and by the measurement of the recoil spectrum. Furthermore, the cross-section behavior close to the production threshold and angular correlations of the decay products in the Higgs-strahlungs-process are sensitive to the spin and quantum numbers and will yield a measurement of the spin and CP properties to a few percent.

\section{Measurements at 500 GeV and up into the multi-TeV region}

Energies of 500 GeV up to 1 TeV and above provide the possibility to directly measure the coupling of the Higgs to the top quark. QCD $t\bar{t}$ bound state effects enhance the cross section at 500 GeV over the simple tree-level expectation shown in Figure \ref{fig:CrossSections} {\it left}, making a measurement possible with polarized beams and integrated luminosities of 1 ab$^{-1}$. At 1 TeV or above the top Yukawa coupling can be measured with sub-5\% accuracy with polarized electron and positron beams. 

The measurement of the tri-linear self-coupling provides direct access to the Higgs potential, and is thus important to establish the Higgs mechanism experimentally. The two double-Higgs production processes, $e^+e^-\rightarrow\mathrm{ZHH}$ and $e^+e^-\rightarrow \mathrm{HH}\nu\nu$, are available for measuring the trilinear self-coupling, with the former reaching its cross-section maximum at a center-of-mass energy of around 500 GeV, while the cross-section for the latter is dominating above $\sim 1$ TeV and increases towards higher energies. Due to the low production cross-sections, the high non-double-Higgs background levels and the complex final state, this is a challenging measurement also at an electron-positron collider. Studies indicate that with integrated luminosities of 2 ab$^{-1}$ it will be possible to provide significant evidence for the self-coupling at 500 GeV, while the same integrated luminosity at 3 TeV will yield a measurement of the self-coupling $\lambda$ on the 12\% level with 80\% polarized electrons. It is expected that a further refinement of the analysis strategies may result in an improvement of these studies in the future. 

In addition to these measurements, which require energies of 500 GeV and above to be accessible, the high luminosity at a multi-TeV linear collider combined with the increasing cross-section of the WW fusion process yields the possibility for precise measurements of Higgs branching ratios extending to rare decays. Accuracies approaching 1\% are expected for the coupling to the $b$ quark, and also the coupling to charm quarks and $\tau$ can be measured with a  sub-2\% accuracy. From the measurement of $\mathrm{H} \rightarrow \mu^+\mu^-$ the coupling to muons is expected with an accuracy on the 7.5\% level. At this energy, the measurement of the ratio of the WW and the ZZ fusion processes also are expected to enables a measurement of the ratio of $g_\mathrm{HWW}/g_\mathrm{HZZ}$ with sub-percent precision.

\section{Detector Requirements from Higgs Physics}

\begin{figure}
\centering
\includegraphics[width=0.85\textwidth]{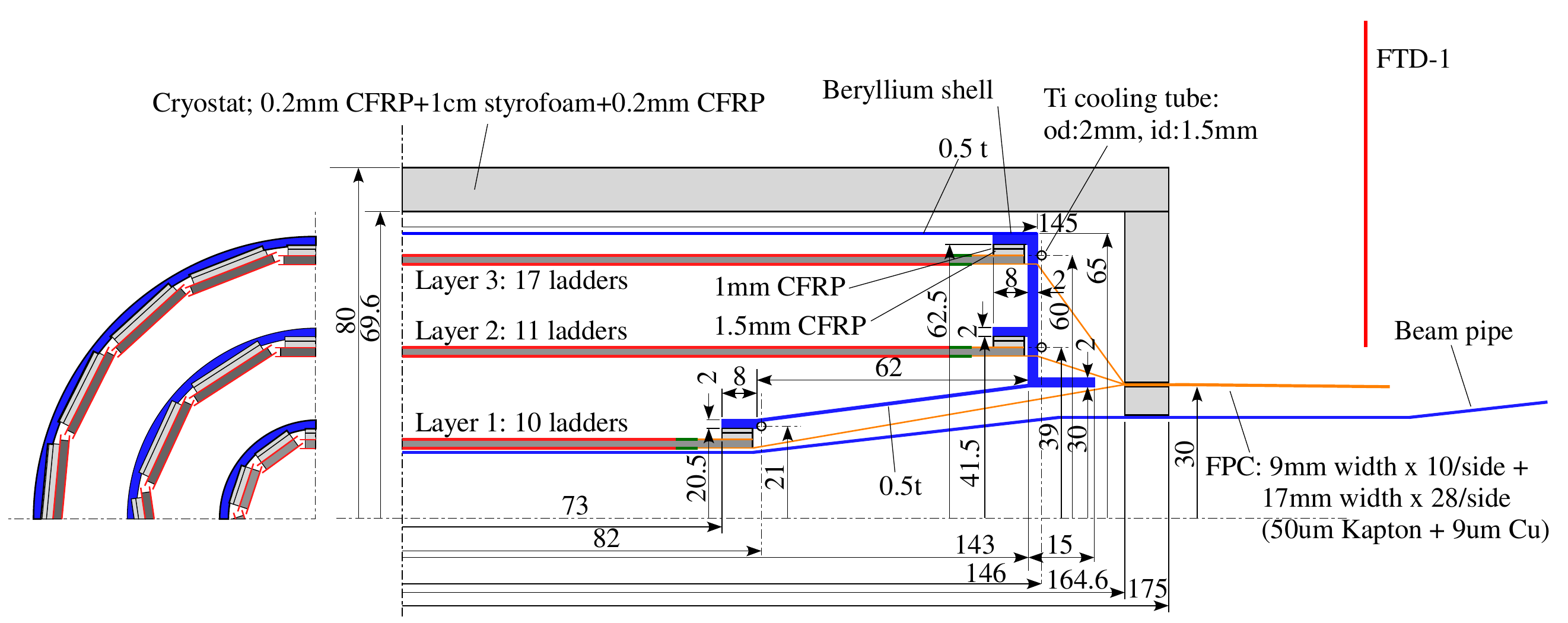}
\caption{Illustration of a possible geometry for the vertex detector of a linear collider detector, here the three double-layer design of ILD. The figure also shows the mechanical support structures, using low-mass materials to reach the smallest possible material budget. Figure taken from \cite{Behnke:2013lya}.}
\label{fig:VTX}Z
\end{figure}

A full exploitation of the possibilities in Higgs physics provided by $e^+e^-$ collisions requires highly performant detector systems. Here, the vertex detector is of particular importance. The measurement of the decays $\mathrm{H} \rightarrow b\bar{b}$,  $\mathrm{H} \rightarrow c\bar{c}$ and  $\mathrm{H} \rightarrow gg$, enabled by the low hadronic background, is only possible with precise secondary vertex reconstruction to distinguish jets originating from $b$ and $c$ quarks as well as light jets from gluons. A highly efficient $b$ jet tagging is also important for the identification of $t\bar{t}\mathrm{H}$ final states for the direct measurement of the top Yukawa coupling and in general for the identification of Higgs bosons in their most probable decay to $b\bar{b}$ in rare final states such as double Higgs production for the measurement of the self coupling.

The requirements of the physics program are met by a vertex detector which has an impact parameter resolution of 
$\sigma_b < 5 \oplus 10/p\,  \mathrm{sin}^{3/2} \theta\, \mu\mathrm{m}$. This results in the requirement of a single point spatial resolution of $\sim$ 5 $\mu$m or better near the interaction point, a very low material budget of a few permille $X_0$ per layer,  pixel occupancies not exceeding a few \% and a first detector layer as close as possible to the interaction point. Several technologies under study have shown or are expected to be capable of reaching these resolution goals while at the same time achieving very low material budgets, among them CMOS active pixel sensors, DEPFET sensors and fine-pitch CCDs as well as hybrid pixel detectors, silicon-on-insulator and 3D-integrated sensors, with more details given in \cite{Behnke:2013lya, Linssen:2012hp, VosVertex}. In particular the latter three options, while still further from demonstrating the required performance or even feasibility, offer the possibility for precise time stamping, which is particularly relevant for a vertex detector for CLIC. 

\begin{figure}
\centering
\includegraphics[width=0.45\textwidth]{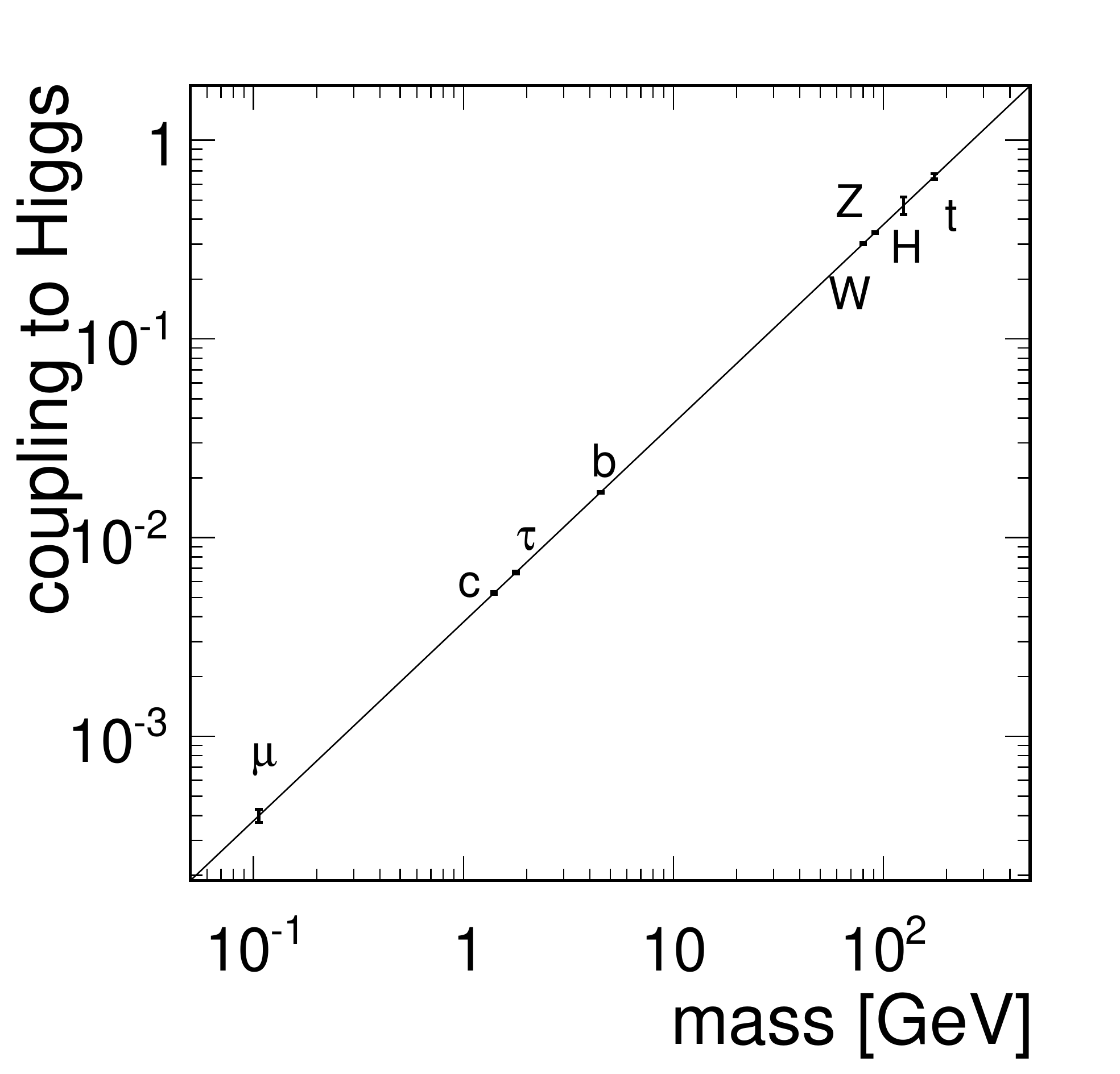}
\hspace{0.05\textwidth}
\includegraphics[width=0.45\textwidth]{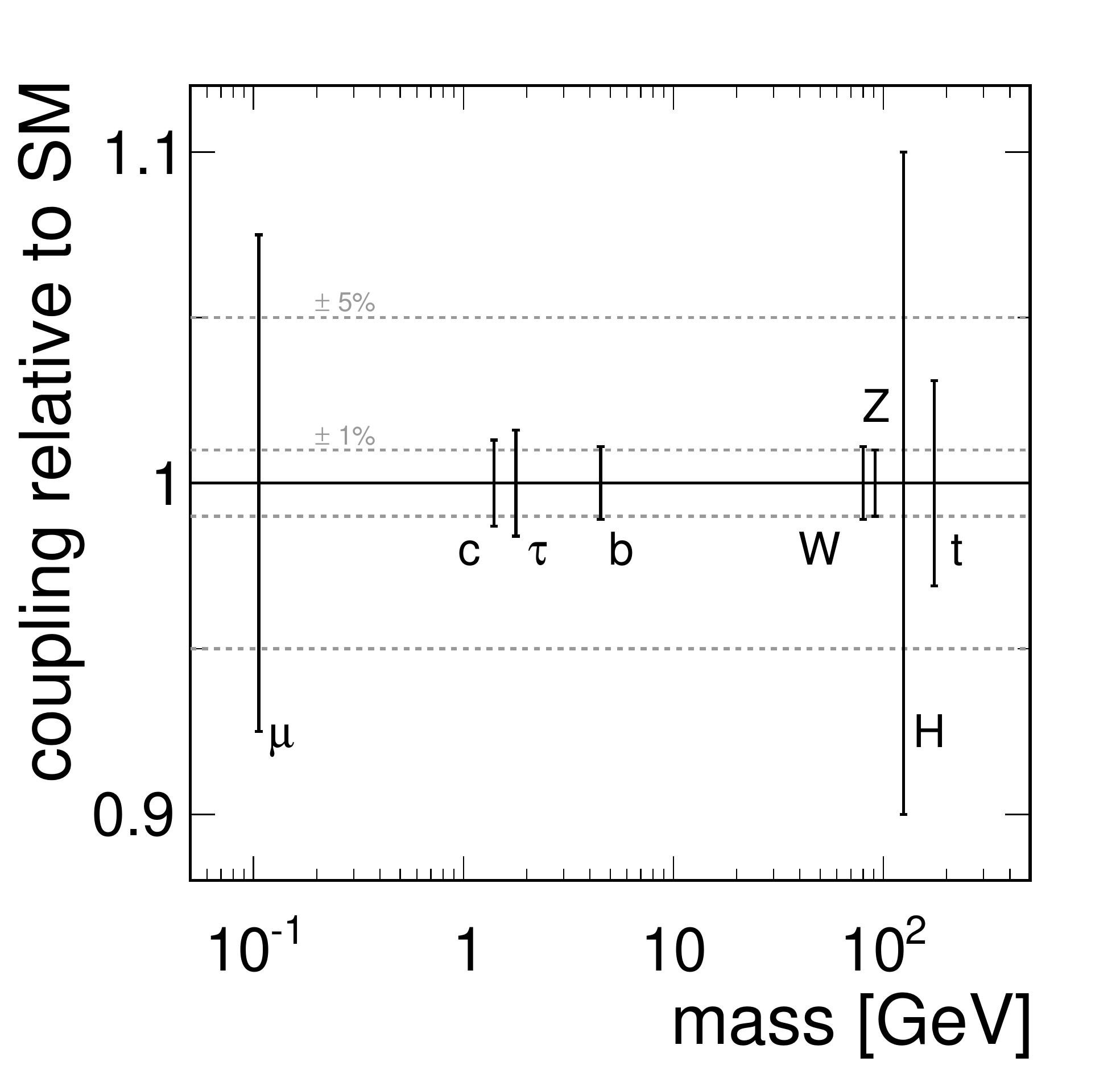}
\caption{Expected precision of model-independently measured couplings to fermions and bosons assuming Standard Model couplings proportional to mass (left) and relative precision of coupling measurements (right) for a full linear collider Higgs physics program extending from 250/350 GeV up to the (multi-) TeV region in several energy stages.}
\label{fig:Couplings}
\end{figure}

Figure \ref{fig:VTX} shows one example for the implementation of a vertex detector at linear colliders, taken from the ILD detector concept. The design consists of three double layers, with the mechanics based on low-mass materials to reach the smallest possible material budget. In addition to a barrel detector, forward tracking is typically provided either by pixel or silicon strip discs. 

Besides these requirements imposed on the vertex detector, there are additional performance goals originating from the Higgs physics program. For the model-independent measurement of the HZ production cross section, and with that of the measurement of the coupling $g_\mathrm{HZZ}$, excellent momentum resolution of the tracking system is required to achieve the highest possible significance in the $\mathrm{Z} \rightarrow \mu^+\mu^-$ and $\mathrm{Z} \rightarrow e^+e^-$ final states. This in turn requires a low material budget all throughout the inner and main tracker regions as well as a main tracker with high spatial resolution. A precise jet energy resolution, provided by particle flow event reconstruction \cite{Thomson:2009rp} and highly granular calorimeters, provides the potential to extend the model-independent ZH cross-section measurements to $\mathrm{Z}\rightarrow q\bar{q}$ decays, and contributes to the explicit reconstruction of Higgs events with hadronic final states. 

The designs of the detector concepts for linear colliders reflect these requirements originating from the Higgs physics program as well as the demands imposed by other physics goals such as the search for and the spectroscopy of New Physics and precision Standard Model measurements. They are met with low-mass, high resolution vertex and main trackers, highly granular electromagnetic and hadronic calorimeters, close-to hermetic coverage also of the forward region and strong solenoidal fields with the magnet coil outside of the calorimeters. Further details on the ILD and SiD concepts for ILC and on the detector concepts for CLIC based on the ILC detectors are given in \cite{Behnke:2013lya, Linssen:2012hp}.

\section{Summary}

Following the discovery of a new boson consistent with the Standard Model Higgs, a linear collider is an excellent option for a comprehensive study of this new form of matter to fully explore the nature of electroweak symmetry breaking. Irrespective of the technology choice for such a collider, its flexibility in energy, further increased by a staged construction, together with the moderate complexity of the final states in $e^+e^-$ collisions, provides the prerequisites for a precise measurement of the properties of this particle substantially beyond the capabilities of the LHC. This includes the measurement of the coupling to fermions and bosons  in a model-independent way, illustrated in Figure \ref{fig:Couplings}, the measurement of mass, spin and CP quantum numbers, and direct access to the Higgs potential through the measurement of the trilinear self-coupling. A comprehensive Higgs physics program at a linear $e^+e^-$ collider spans a wide energy range, making full use of the capabilities of the accelerator. Such a program also imposes strict requirements on the detector systems. Of particular importance here is the vertex detector, which has to be capable to separate $b$, $c$ and light jets to enable a comprehensive measurement of the Higgs couplings to different fermions, and to allow the identification of complex, low-cross-section final states. In addition, high-resolution tracking for charged leptons is mandatory to achieve the best possible model-independent measurement of the coupling to the Z boson, which is further supported by an excellent jet energy resolution.

\end{document}